\documentclass[a4paper,12pt]{amsart}

\usepackage[colorlinks=true]{hyperref}
\hypersetup{urlcolor=blue, citecolor=blue, linkcolor=blue}

\usepackage{amsmath,siunitx,dsfont,amssymb,bm,enumerate,mathrsfs, physics,bbold}

\usepackage{graphicx}

\usepackage{mathtools}
\mathtoolsset{showonlyrefs}

\newcommand{\oa}{\mathscr{O}}
\newcommand{\ob}{\mathscr{O}^{\prime}}

\newcommand{\fa}{\Omega}
\newcommand{\fb}{\Omega^{\prime}}
\renewcommand{\t}{\tau}
\newcommand{\x}{\xi}
\newcommand{\y}{\upsilon}
\newcommand{\z}{\zeta}

\newcommand{\ab}{{\Phi^{\prime}}}
\newcommand{\tb}{\Theta}
\newcommand{\xb}{\Xi}
\newcommand{\yb}{\Upsilon}
\newcommand{\zb}{\mathcal{Z}}

\newcommand{\ba}{{\Phi}}
\newcommand{\ta}{\mathrm{T}}
\newcommand{\xa}{\mathrm{X}}
\newcommand{\ya}{\mathrm{Y}}
\newcommand{\za}{\mathrm{Z}}

\newcommand{\rv}{{v}}

\newcommand{\bv}{\mathbf{v}}

\newcommand{\M}[1]{\begin{bmatrix} #1 \end{bmatrix}}

\newcommand{\ii}{\mathds{1}}
\newcommand{\RN}{\mathds{R}}

\DeclareMathOperator{\diag}{diag}

\DeclareMathOperator{\id}{id}

\newcommand{\p}[1]{\partial_{#1}}

\newcommand{\B}[1]{\mathbf{#1}}

\newcommand{\ps}[1]{\sigma_{#1}}

\theoremstyle{remark}

\begin{document}
\begin{abstract}
The relative motion of material, point-like observers is analysed in terms of coordinate maps between the respective rest-frame of each observer. Under the assumption these maps are $C^{2}$-regular, conservation laws are deduced, which in turn are found to yield wave equations familiar as scalar and spinor fields. Wave-particle dualism is then understood as a property of these coordinate maps as opposed to an inherent property of the point-like observers themselves. Additionally it is shown that the relative acceleration of these observers can only be realised in this context by the introduction of additional potential fields.  It is shown that if each observer is to be associated with an inertial frame of reference, then the potential field is necessarily a massless gauge field.
\end{abstract}

\title[Waves, particles and coordinate transformations]{On the intrinsic wave-particle duality of coordinate maps}
\author{Tony Lyons}
\address{Department of Computing \& Mathematics, South East Technological University, Waterford, Ireland.}
\email{tony.lyons@setu.ie}
\maketitle{}
\tableofcontents
\section{Introduction}
The wave-particle duality associated with massive particles, first postulated by de Broglie in his seminal work \cite{deB1923, deB1925} has provided the theoretical basis for the later development of wave mechanics (cf. \cite{Sch1926, Sch2003c}) and its attendant successes in describing a vast array of microscopic phenomena. Nevertheless, the interpretation of the wave-particle duality has remained somewhat ambiguous since its introduction, in the sense that a material point particle must be associated with an extended wave-like character. While the theoretical interpretation of the wave-particle duality has remained a vexing issue, its experimental confirmation was realised almost immediately for electrons by Davisson \& Germer (cf. \cite{DG1927}). Wave-particle duality  is now routinely confirmed for more substantial material bodies, see for instance \cite{Arnetal1999} which experimentally confirms  wave-particle duality for $\mathrm{C_{60}}$ molecules, while \cite{Getal2011,PMWB2019,Setal2008,Setal2020,TM2021} provide further confirmation of wave-particle dualism for matter. The aim of the current work is to investigate the wave-particle duality from the perspective of relational mechanics, whereby all motion is treated in strictly relative terms. It is demonstrated that a consistent application of the relativity principle (see \cite{Mol1972} for instance) suggests an apparent contradiction with the exclusively point-like interpretation of massive particles.

The interpretation of wave-particle dualism as an inherent feature of space-time coordinate transformations is motivated by the assumption that all observers are associated with a rest-frame whose coordinates provide a valid frame work for the description of physical laws. The maps between these reference frames are considered as maps with $C^{2}$-regularity with respect to the space-time coordinates of their domain. This regularity is found to ensure the existence of conservation laws for the energy-momentum of each material particle when observed from any frame of reference. Moreover, a consistent implementation of the nonlinear relativistic energy-momentum constraint combined with this $C^{2}$-regularity is also found to necessitate the introduction of appropriate potentials to describe the case of non-uniform relative motion between observers and their respective rest-frames.

A reformulation of the conservation laws associated with the $C^{2}$-regularity of the coordinate maps and the energy-momentum constraint is found to reproduce the Klein-Gordon wave equation. This wave-equation is found to be an intrinsic property of the space-time mappings as opposed to a feature of the material bodies themselves. It is noted that this is entirely consistent with Wigner's observation that the fundamental particles of nature must belong to some representation of the Lorentz group (cf. \cite{Wig1939}). The intrinsic wave character of the coordinate mapping is found to reproduce the Dirac equation (cf. \cite{Dir1928, Dir1981}) as a reformulation of the inverse function theorem associated with the coordinate transformation. Finally, the case of observers in a state of relative acceleration is found to necessitate the introduction of potential fields, familiar as massless gauge fields.  Thus it appears wave-particle dualism  is essentially a physical manifestation of the relativity principle for interacting observers.

\section{Relative motion and coordinate transformations}
Consider the case of  two observers $\oa$ and $\ob$, and associated with each observer are their respective rest frames $\fa$ and $\fb$.
The location of $\oa$ with reference to $\fa$ is always of the form $(t,0,0,0)$ for $t\in\RN$ while the location of $\ob$ with reference to this frame is of the form $(t,x,y,z)\in\RN^4$, where $x$, $y$ and $z$ may be functions of $t$, depending on the state of relative motion between the two observers.  Conversely, in the frame $\fb$ the space-time coordinates of the observer $\ob$ are always of the form $(\t,0,0,0)$ with $\t\in\RN$, while the coordinates of $\oa$ are now of the form $(\t,\x,\y,\z)$, where again $\x$, $\y$ and $\z$ may be functions of $\t$ depending on the state of relative motion between the two observers.

It is clear that to every space-time location in $\fa$ there are corresponding space-time coordinates in the frame $\fb$, and we say $(t,x,y,z)\sim(\t,\x,\y,\z)$ if these coordinates represent the same event as observed in the frames $\fa$ and $\fb$ respectively. Since every coordinate in $\fa$ has a counterpart in $\fb$ it follows that there exist maps from one frame to the other of the form
\begin{equation}
\label{eq:ab}
\begin{aligned}
\ab&:\fa \rightarrow \fb
\\
\ab&:\left(t , x , y , z\right) \mapsto \left(\tb , \xb , \yb , \zb\right)\big{|}_{(t , x , y , z)} =  \left(\t , \x , \y , \z\right)
\end{aligned}
\end{equation}
with the inverse map of the form
\begin{equation}
\label{eq:ba}
\begin{aligned}
\ba &: \fb \rightarrow \fa
\\
\ba &: \left(\t , \x , \y , \z\right) \mapsto \left(\ta , \xa , \ya , \za\right)\big{|}_{(\t , \x , \y , \z)} = \left(t , x , y , z\right).
\end{aligned}
\end{equation}
In the following it shall be assumed that  $\ab\in C^{2}\left( \fa \right)$ and $\ba\in C^{2}\left( \fb \right)$.  Moreover, given $$(\ba\circ\ab)(t,x,y,z)=(t,x,y,z)\quad (\ab\circ\ba)(\t,\x,\y,\z) = (\t,\x,\y,\z)$$ then the aforementioned regularity requirements ensures the inverse function theorem applies (see \cite{Rud1976} for instance):
\begin{equation}\label{eq:IFT}
 \M{\tb_{t} & \xb_{t} & \yb_{t} & \zb_{t} \\ \tb_{x} & \xb_{x} & \yb_{x} & \zb_{x} \\ \tb_{y} & \xb_{y} & \yb_{y} & \zb_{y} \\ \tb_{z} & \xb_{z} & \yb_{z} & \zb_{z}}
 \M{\ta_{\t} & \xa_{\t} & \ya_{\t} & \za_{\t} \\ \ta_{\x} & \xa_{\x} & \ya_{\x} & \za_{\x} \\ \ta_{\y} & \xa_{\y} & \ya_{\y} & \za_{\y} \\ \ta_{\z} & \xa_{\z} & \ya_{\z} & \za_{\z}}
 =
 \M{1 & 0 & 0 & 0 \\ 0 & 1 & 0 & 0 \\ 0 & 0 & 1 & 0 \\ 0 & 0 & 0 & 1},
\end{equation}
where subscripts denote partial differentiation with respect to the corresponding variables.  In the following it shall be assumed that the $(x,y,z)$-axes of $\fa$ align with the corresponding $(\x , \y , \z)$-axes of $\fb$, with no rotation of one frame relative to another taking place throughout the course of their relative  motion. To simplify the following analysis, it is assumed the motion of $\ob$ relative to $\fa$ takes place along the $x$-axis of $\fa$, and as such it shall only be necessary to consider the $(t,x)$ coordinates of $\ob$ in this frame, in most cases. Likewise, it is assumed the motion of $\oa$ relative to the frame $\fb$ shall take place along the $\x$-axis of this reference frame and so it is again only necessary to refer to the $(\t,\x)$-coordinates of $\oa$ in $\fb$ for the most part.

\subsection{Local Lorentz transformations}\label{sec:Local}
The principle of general relativity implies the coordinate systems $\fa$ and $\fb$ should be treated equivalently from a physical point of view (see \cite{Mol1972} for further discussion of this principle as it relates to the description of acceleration in relative terms). In this regard the relative motion between the pair may be described in the same fashion in the reference frame $\fa$ or $\fb$. At some instant the observer $\ob$ may occupy the space-time location $(t,x,y,z)\in\fa$ and is moving with some velocity $\bv$ relative to $\oa$ (and thus $\fa$), while at the same instant the observer $\oa$ occupies some space-time location $(\t , \x , \y , \z) \in \fb$
and is moving with velocity $-\bv$. The relative motion of $\oa$ and $\ob$ is described as uniform relative motion if $\bv$ is constant throughout $\fa$ and $\fb$, otherwise, the relative velocity $\bv$ has instantaneous values at various stages of the relative motion. However it is always the case that if a point $(\t,\x,\y,\z)\in\fb$ moves with velocity $\bv$ relative to $\fa$ then the corresponding point $(t,x,y,z)\in\fa$ is found to move with velocity $-\bv$ relative to the frame $\fb$, instantaneously.

Since we consider $\oa$ and $\ob$ to be in relative motion along the respective $x$  and $\x$ axes of $\fa$ and $\fb$, the relative velocity of $\ob$ with reference to $\fa$ is simply of the form $\bv=(\rv,0,0)$, and  it is convenient for now to consider space-time locations of $\ob$ with reference to $\fa$ using only their $(t,x)$-coordinates. If $v$ is the instantaneous velocity of $\ob$ with reference to $\fa$, we consider a pair of space-time locations $(t,x)\in\fa$ and $(t+dt,x+dx)\in\fa$, where the coordinate displacements $(dt,dx)$ have arbitrarily small magnitude. The coordinate map \eqref{eq:ab} now gives
\begin{equation}\label{eq:dxb}
 d\t = \tb_{t}dt + \tb_{x}dx \qquad d\x=\xb_{t}dt+\xb_{x}dx.
\end{equation}
As such the instantaneous velocity along the trajectory connecting these space-time locations in each frame are related according to
\begin{equation}\label{eq:vavb}
 \frac{d\x}{d\t} = \frac{ \xb_{t}+\xb_{x}\frac{dx}{dt} }{\tb_{t}+\tb_{x}\frac{dx}{dt} }.
\end{equation}
If the coordinate displacements $(dt,dx)$  are related according to $dx=\rv dt$, then this corresponds to the instantaneous trajectory of $\ob$ with reference to $\fa$ and as such it must be the case that
\begin{equation}\label{eq:dxdt=v}
 0=\frac{d\x}{d\t}=\frac{ \xb_{t}+\xb_{x}\rv}{\tb_{t}+\tb_{x}\rv} \implies \xb_{t} = -\rv\xb_{x}.
\end{equation}
Alternatively, if the velocity along the trajectory vanishes with reference to $\fa$, that is $dx=0$, then the velocity along the trajectory with reference to $\fb$ must be $-\rv$ and so it follows that
\begin{equation}\label{eq:dxdt=0}
 -\rv=\frac{d\x}{d\t}=\frac{\xb_{t}}{\tb_{t}} \implies \xb_{t} = -\rv\tb_{t}.
\end{equation}
Thus equations \eqref{eq:dxdt=v}--\eqref{eq:dxdt=0} ensure that
\begin{equation}\label{eq:ttxx}
 \tb_{t}=\xb_{x}.
\end{equation}

The Lorentz transformation between reference frames in a state of uniform relative motion follows from the constancy of the speed of light $c\approx\SI{2.998e8}{\meter\per\second}$ in all frames of reference. Specifically, if $(dt,dx)$ are coordinate displacements in $\fa$ related by $dx=cdt$ then the associated coordinate displacements in the frame $\fb$ are also related by $d\x=cd\t$. As such, equations \eqref{eq:vavb}--\eqref{eq:ttxx} now give
\begin{equation}\label{eq:dxdt=c}
 c=\frac{d\x}{d\t} = \frac{\xb_{t} + \xb_{x}c}{\tb_{t} + \tb_{x}c} \implies \tb_{x} = -\frac{\rv}{c^2}\tb_{t}.
\end{equation}
In line with the principle of relativity, the components $\{\ta,\xa\}$ of the inverse map $\ba$ (cf. equation \eqref{eq:ba}) are related by
\begin{equation}
\ta_{\t}=\xa_{\x}\quad \xa_{\t}=\rv\ta_{\t}\quad \ta_{\x}=\frac{\rv}{c^2}\ta_{\t}.
\end{equation}
As such, the inverse function theorem now requires
\begin{equation}
\M{\ta_{\t} & \rv\ta_{\t} \\ \frac{\rv}{c^2}\ta_{\t} & \ta_{\t}}\M{\tb_{t} & -\rv\tb_{t} \\ -\frac{\rv}{c^2}\tb_{t} & \tb_{t}} = \M{1 & 0 \\ 0 & 1}
\end{equation}
from which we deduce
\begin{equation}\label{eq:tttt}
\ta_{\t}\tb_{t} = \frac{1}{1-\frac{\rv^2}{c^2}}.
\end{equation}
Since the map $\ta$ is expected to be obtained from the map $\tb$ under a reverse of velocity, namely $\rv\to-\rv$, from the constraint \eqref{eq:tttt} we  expect $\ta_{\t}=\tb_{t}=\pm\frac{1}{\sqrt{1-\frac{\rv^2}{c^2}}}$, which also ensures $\ta_{t}$ and $\tb_{\t}$ are independent of the direction of $v$.  Choosing for the moment the positive solution, the Jacobian of the coordinate map is given by
\begin{equation}\label{eq:J=1}
 J=\tb_{t}^2-c^2\tb_{x}^2 = \xb_{x}^2-\frac{1}{c^2}\xb^2_{t}= 1,
\end{equation}
and so the coordinate displacements $(dt,dx)$ and $(d\t,d\x)$ are related according to
\begin{equation}\label{eq:Lorentz}
 d\t = \frac{dt-\frac{\rv}{c^2}dx}{\sqrt{1-\frac{\rv^2}{c^2}}} \quad d\x = \frac{dx-\rv dt}{\sqrt{1-\frac{\rv^2}{c^2}}},
\end{equation}
which is simply a localised form of the familiar Lorentz transformation.

If $(mc^2,0,0,0)$ is the constant energy-momentum vector of $\ob$ with reference to its rest-frame $\fb$, then its instantaneous energy-momentum with reference to the frame $\fa$ is given by
\begin{equation}
 \left(E,p,0,0\right)  = \left(\frac{mc^2}{\sqrt{1-\frac{\rv^2}{c^2}}} , \frac{m\rv}{\sqrt{1-\frac{\rv^2}{c^2}}},0,0\right),
\end{equation}
from which one easily deduces the relativistic energy-momentum constraint
\begin{equation}\label{eq:enrmom}
 E^2- p^2c^2=m^2c^4.
\end{equation}
Moreover, the local Lorentz transformation may now be written in terms of the instantaneous energy-momentum vector of $\ob$ with reference to $\fa$ according to
\begin{equation}\label{eq:dt&dx}
 d\t = \frac{1}{mc^2}\left(Edt-pdx\right) \quad d\x = \frac{1}{mc^2}\left(Edx-pc^2dt\right),
\end{equation}
which will prove more convenient in what follows. Moreover it is clear that the energy-momentum of $\ob$ with reference to $\fa$ can be characterised by
\begin{equation}
 E=mc^2\tb_{t} \quad p = -mc^2\tb_{x},
\end{equation}
while the energy momentum condition \eqref{eq:enrmom} is simply a reformulation of the condition $J=1$ (cf. equation \eqref{eq:J=1}).

\subsection{Regularity \& conservation laws}\label{sec:Regularity}
 Given the regularity conditions $\ab\in C^{2}(\fa)$ and $\ba\in C^{2}(\fb)$, Schwartz's theorem (see \cite{Die1960} for instance)  implies $\ab_{tx} = \ab_{xt}$ throughout $\fa$. A state of relative motion between $\oa$ and $\ob$ as characterised by the coordinate transformation $\ab$ requires
\begin{equation}\label{eq:MEp}
\M{\tb_{t} & \xb_{t} \\ \tb_{x} & \xb_{x}} = \frac{1}{mc^2}\M{E & -pc^2 \\ -p & E}
\end{equation}
whose inverse
\begin{equation}\label{eq:MEp'}
\M{\ta_{\t} & \xa_{\t} \\ \ta_{\x} & \xa_{\x}} = \frac{1}{mc^2}\M{E & pc^2 \\ p & E}
\end{equation}
is easily deduced from the inverse function theorem (cf. equation \eqref{eq:IFT}). The $C^{2}$-regularity of $\tb$ now ensures that
\begin{equation}\label{eq:force}
 \tb_{tx} = \tb_{xt} \implies E_{x} + p_{t} = 0,
\end{equation}
which simply relates the acceleration to the gradient of the energy in the vicinity of $\ob$ with reference to the frame $\fa$.
Similarly the $C^{2}-$regularity of $\xb$ also implies
\begin{equation}\label{eq:conserve}
 \xb_{xt} = \xb_{tx} \implies \frac{1}{c^2}E_{t} + p_{x} = 0,
\end{equation}
which ensures the conservation of energy-momentum of $\ob$ along its trajectory through the frame $\fa$.

Moreover, it follows that equation \eqref{eq:MEp} may also be written according to
\begin{equation}\label{eq:ThEp}
 \M{\tb_{t} & c^2\tb_{x} \\ \tb_{x} & \tb_{t}} = \M{\xb_{x} & -\xb_{t} \\ -\frac{1}{c^2}\xb_{t} & \xb_{x} }= \frac{1}{mc^2}\M{E & -pc^2 \\ -p & E},
\end{equation}
which along with the conditions \eqref{eq:force}--\eqref{eq:conserve} now yield the following wave equations:
\begin{subequations}
\begin{align}
\label{eq:wt} \tb_{tt} - c^2\tb_{xx} &= 0\\
\label{eq:wx} \xb_{tt} - c^2\xb_{xx} &= 0.
\end{align}
\end{subequations}
The wave equations \eqref{eq:wt}--\eqref{eq:wx} must be considered in conjunction with the nonlinear constraints \eqref{eq:J=1} following from the relativistic energy-momentum condition.

\subsection{Coordinate maps as scalar fields}\label{sec:scalar}
The constraint that must be satisfied by the function $\tb$ is \eqref{eq:J=1} which is equivalent to the mass-energy constraint:
\[ \tb_{t}^{2}-c^2\tb_{x}^{2} = 1 \iff E^2 - p^2c^2=m^2c^4.\]
In addition the map must satisfy the wave equation \eqref{eq:wt} which is simply a reformulation of the conservation of energy-momentum
\[\tb_{tt} - c^2\tb_{xx} = 0 \iff \frac{1}{c^2}E_{t}+p_{x} = 0\]
which is itself a consequence of the $C^{2}$--regularity of the maps $\tb$ and $\xb$.

We introduce the map $\psi(t,x)\in C^{2}(\fa)$ defined according to
\begin{equation}\label{eq:psi0}
 \tb(t,x)=\rho(\psi(t,x))\in C^{2}(\fa).
\end{equation}
where $\rho(\cdot)$ and $\psi(\cdot,\cdot)$ are as yet otherwise arbitrary.
The condition \eqref{eq:J=1} may now be reformulated according to
\begin{equation}\label{eq:rho'}
\rho'(\psi)^2\left[\psi_{t}^2 - c^2\psi_{x}^2\right] = 1,
\end{equation}
in which case $\rho'(\psi)$ and $\psi_{t}^2 - c^2\psi_{x}^2$ are both non-zero in $\fa$.
Meanwhile the conservation equation \eqref{eq:wt} may be written according to
\begin{equation}\label{eq:rho''}
 \rho''(\psi)\left[\psi_{t}^2 - c^2\psi_{x}^2\right] + \rho'(\psi)\left[\psi_{tt} - c^2\psi_{xx}\right] = 0,
\end{equation}
Physically, the conditions
\[E=mc^2\rho^{\prime}(\psi)\psi_{t}\quad p=-mc^2\rho^{\prime}(\psi)\psi_{x}\]
with $E$ and $p$ both non-zero (since $\oa$ and $\ob$ are in a state of relative motion) also suggest $\psi_{t}$ and $\psi_{x}$ are both non-zero in $\fa$.

The constraint \eqref{eq:rho'} may be used to reformulate \eqref{eq:rho''} according to
\begin{equation}\label{eq:wave1}
 \frac{\rho''(\psi)}{\rho'(\psi)^3}+\psi_{tt} - c^2\psi_{xx} = 0,
\end{equation}
while upon multiplying equation \eqref{eq:wave1} by $\psi_{t}$ one obtains
\begin{equation}\label{eq:new_rho''}
 \frac{1}{2}\partial_{t}\left[-\frac{1}{\rho'(\psi)^2}+\psi_{t}^{2}\right] - c^2\psi_{xx}\psi_{t} = 0.
\end{equation}
Equation \eqref{eq:rho'} now allows one to substitute for $\frac{1}{\rho'(\psi)^2}$, which then ensures
\begin{equation}\label{eq:lnx}
 \psi_{xt}\psi_{x} - \psi_{xx}\psi_{t} = 0 \implies \left(\frac{\psi_{x}}{\psi_{t}}\right)_{x} = 0.
\end{equation}
In a similar manner, multiplying equation \eqref{eq:wave1} by $\psi_{x}$ also ensures
\begin{equation}\label{eq:lnt}
 \left(\frac{\psi_{x}}{\psi_{t}}\right)_{t} = 0,
\end{equation}
in which case $\frac{\psi_{x}}{\psi_{t}}$ is a constant. It follows that $\psi(t,x)$ is in general of the form
\begin{equation}\label{eq:psiwk}
 \psi(t,x) = \psi(\omega t- k x) \equiv \psi(s),
\end{equation}
where $\omega$ and $k$ are constants.

Rewriting equation \eqref{eq:rho'} in terms of $\psi(s)$, it follows at once that
\begin{equation}\label{eq:rho'psi'}
 \omega_{0}^{2}\rho'(\psi)^2\psi'(s)^2 = 1,
\end{equation}
where $\omega_{0}^2=\omega^2-k^2c^2\neq0$.
Moreover, the inverse function theorem (cf. \cite{Rud1976}) may be applied to equation \eqref{eq:rho'psi'} thus ensuring
\begin{equation}
 \pm\omega_{0}(\rho\circ\psi)(s)  = \id(s) = \omega t - kx
\end{equation}
in which case $\tb(t,x)$ is a linear map, specifically
\begin{equation}\label{eq:twk}
 \tb(t,x) = \pm \frac{\omega t - kx}{\omega_{0}}.
\end{equation}
Taking the negative sign in equation \eqref{eq:twk} means events in the frame $\fa$ are labelled in reverse chronological order compared to their labelling in $\fb$. In the current context, the principle of relativity suggests there is no absolute sense as to which coordinate frame uses the truly ``forward'' progressing time coordinate.

Note that $\psi(t,x)$ as defined by equation \eqref{eq:psi0} is quite arbitrary. In particular if given two maps $\tb_{1}(t,x)=\rho(\phi(t,x))$ and $\tb_{2}(t,x)=\rho(\psi(t,x))$ with  $\phi= \kappa \psi$ for some non-zero constant $\kappa$, then equation \eqref{eq:rho'} must be satisfied in each case, and as such
\begin{equation}
\rho'(\kappa\psi)^2\kappa^{2}\left[\psi_{t}^2 - c^2\psi_{x}^2\right] = \rho'(\psi)^2\left[\psi_{t}^2 - c^2\psi_{x}^2\right] =1.
\end{equation}
Thus it appears
\begin{equation}
 \kappa^{2}\rho^{\prime}(\kappa\psi)^{2} = \rho^{\prime}(\psi)^{2}
\end{equation}
from which it must follow
\begin{equation}\label{eq:log'}
 \rho'(\psi) \simeq \frac{1}{\psi},
\end{equation}
up to some multiplicative constant.
It follows that the map $\tb(t,x)$ may be generally defined according to
\begin{equation}\label{eq:psitau}
\tb(t,x) = \frac{\mu}{mc^2}\ln\psi(\omega t - kx),
\end{equation}
where $\mu$ is a constant whose unit must clearly be that of action, that is  $[\mu]=\SI{}{J\cdot s}$.
The energy-momentum as given by $E=mc^2\tb_{t}$ and $p=-mc^2\tb_{x}$ are equivalently written according to the eigenvalue problem
\begin{equation}\label{eq:eigenprob}
 \mu\psi_{t} = E\psi \qquad -\mu\psi_{x} = p\psi,
\end{equation}
and equation \eqref{eq:rho''} may be reformulated according to
\begin{equation}\label{eq:wpsi}
\frac{1}{c^2}\psi_{tt} - \psi_{xx} - \frac{m^2c^2}{\mu^2}\psi = 0.
\end{equation}
While it is always desirable to deduce all conditions as far as possible from first principles, it appears that in the current framework one must actively choose the value of $\mu$, specifically whether $\mu$ should be chosen as a real or imaginary constant.

The possible choice $\mu = \mu_{1}+i\mu_{2}$ with $\mu_{n}\in\RN$ for $n\in\{1,2\}$, provides no greater generality since $\tb(t,x)$ is expected to be real valued. If $\mu$ is a general complex constant, then it must be the case that $\ln\psi$ is a complex valued function, with
\[\Re{\ln\psi(t,x)} =a_{1}(t,x),  \quad \Im{\ln\psi(t,x)} =a_{2}(t,x).\]
The condition $\tb(t,x)$ is real valued now ensures
\begin{equation}
 mc^2\tb = \mu_{1}a_{1}-\mu_{2}a_{2} + i\left(\mu_{1}a_{2}+\mu_{2}a_{1}\right),
\end{equation}
is also real valued and as such it is clear that
\[a_{1}(t,x) = -\frac{\mu_{1}}{\mu_{2}}a_{2}(t,x). \]
It follows that $\tb(t,x)$ may be written exclusively in terms of $a_{1}(t,x)$ or $a_{2}(t,x)$ according to
\[\tb(t,x) = \frac{\mu_{1}^2+\mu_{2}^2}{\mu_{1}mc^2}a_{1}(t,x) = -\frac{\mu_{1}^2+\mu_{2}^2}{\mu_{2}mc^2}a_{2}(t,x).\]
Thus the possible values of $\mu$ to be considered are either real or imaginary values, without loss of generality.

It is well established that $\mu=\frac{\hbar}{i}$ is consistent with the Hamilton-Jacobi equations of classical mechanics (cf. \cite{MS1964}). Moreover it is noteworthy that once $\mu$ is imaginary the wave-equation \eqref{eq:wpsi} has a well-posed Cauchy problem (cf. \cite{Wal1984}), while its solution space is naturally endowed with an inner-product structure and an associated orthogonal basis, as well as a Hilbert space structure under appropriate conditions (cf. \cite{Bre2011}). In particular, the $\tau$-difference between events in $\fa$ of the form $(t,x_{1})$ and $(t,x_{2})$ is given by
\[\int_{x_{1}}^{x_{2}}\tb_{x}(t,x)dx=\int_{x_{1}}^{x_{2}}\frac{1}{\psi}\psi_{x}dx = \int_{x_{1}}^{x_{2}}e^{-\frac{1}{\mu}(Et-px)}\p{x}e^{\frac{1}{\mu}(Et-px)}dx.\]
On the other hand, a general solution of \eqref{eq:wpsi} takes the form
\[\psi(t,x) = \sum_{n=1}^{N}a_{n}e^{\frac{1}{\mu}(E_{n}t-p_{n}x)},\]
while the observer $\ob$ associated $\psi(t,x)$ has a unique energy-momentum once
\[\int_{x_{1}}^{x_{2}}\tb_{x}(t,x)dx =  \int_{x_{1}}^{x_{2}}e^{-\frac{1}{\mu}(Et-px)}\p{x}\psi(t,x)dx,\]
is time-independent. This is possible only if  $e^{-\frac{1}{\mu}(Et-px)}$ form an orthogonal set of basis functions, thus implying $\mu$ is imaginary.
 Moreover, there is significant experimental evidence to support wave-particle duality for massive particles in nature, something which appears to be realised only when $\mu$ is imaginary. Thus a clear choice for this constant is $\mu=\frac{\hbar}{i}$ where $\hbar = \SI{1.05457e-34}{J\cdot s}$ is the reduced  Planck constant. The eigenvalue problem \eqref{eq:eigenprob} now becomes
\begin{equation}\label{eq:QM}
 -i\hbar\psi_{t} = E\psi \qquad i\hbar\psi_{x} = p\psi.
\end{equation}
while equation \eqref{eq:wpsi} now yields
\begin{equation}\label{eq:KG}
 \frac{1}{c^2}\psi_{tt} - \psi_{xx} + \frac{m^2c^2}{\hbar^2}\psi = 0.
\end{equation}

A general coordinate transformation as given by \eqref{eq:IFT} may always be reduced to the form \eqref{eq:dxb} by means of appropriate rotations (cf. \cite{Wig1939}). Hence equations \eqref{eq:QM}--\eqref{eq:KG} are generalised in a straightforward manner to relative motion in an arbitrary direction. In particular, it may be shown that the coordinate map $\tb(t,x,y,z)$ may always be written in the form
\begin{equation}\label{eq:tau3}
\tb(t,x,y,z) =- \frac{i\hbar}{mc^2}\ln\psi(t,x,y,z),
\end{equation}
with
\begin{equation}
-i\hbar\psi_{t} = E\psi \quad i\hbar\nabla\psi=\B{p}\psi,
\end{equation}
where $\B{p}=(p_{1},p_{2},p_{3})$ are the momentum components of $\ob$ with reference to the frame $\fa$. The corresponding wave equation is then simply
\begin{equation}\label{eq:KG3}
 \frac{1}{c^2}\psi_{tt} - \nabla^2\psi +\frac{m^2c^2}{\hbar^2}\psi = 0,
\end{equation}
the familiar Klein-Gordon wave equation.

\subsection{Commutators}
While our analysis so far has related to maps $\psi(t,x)=e^{\frac{i}{\hbar}(Et-px)}$ associated with coordinate transformations characterised by unique energy momentum $(E,p)$, we note that the solution space of \eqref{eq:KG} is linear, and as such any linear combination of such wave-functions is also an admissible solution. In particular, it means one may construct generalised coordinate maps $\psi$ wherein the energy-momentum is associated with an \emph{ensemble of possible values}, something characteristic of quantum systems.

Moreover, as demonstrated in \cite{MS1964}, the classical commutativity of position and momentum as give by $px-xp=0$ is realised if and only if $\left(\hat{p}\hat{x}-\hat{x}\hat{p}\right)\psi=i\hbar\psi$ where $\hat{x}\psi\equiv x\psi$ and $\hat{p}\psi \equiv i\hbar\p{x}\psi$. Equation \eqref{eq:QM} means we may obtain the classical (i.e. measured) momentum of $\ob$ with respect to $\fa$ according to $p=i\hbar\frac{\psi_{x}}{\psi}$. Classical commutativity of position and momentum now requires
\begin{equation}
 px-xp = 0 \implies i\hbar\frac{\psi_{x}}{\psi}x - i\hbar x\frac{\psi_{x}}{\psi}=0.
\end{equation}
As such we have
\begin{equation}
 i\hbar\psi_{x}x-i\hbar x\psi_{x} = 0 \implies i\hbar\p{x}(x\psi) - i\hbar\psi - i\hbar x\psi_{x}=0,
\end{equation}
and so it follows that
\begin{equation}
 \left[i\hbar\p{x},x\right]\psi = i\hbar\psi \implies [\hat{p},\hat{x}]\psi=i\hbar\psi,
\end{equation}
where $[\cdot,\cdot]$ denotes the usual operator commutator. Thus we conclude that the position and momentum of $\ob$ are not simultaneously definable with reference to $\fa$ using $\psi(t,x)$. This is not to say $\ob$ is not a localised particle, in fact $\ob$ always has a well-defined location $(\tau,0)\in\fb$ for all $\tau\in\RN$. Instead it means the mapping $\ab:\fa\to\fb$ or its inverse $\ba:\fb\to\fa$ cannot fully localise $(\tau,0)\in\fb$ with reference to $\fa$. However, this does not contradict any of our earlier observations.

The derivation of the local Lorentz transformation in \S \ref{sec:Local} relied on specifying the velocity of $\ob$ with reference to $\fa$, which means the ratio $v=\frac{pc^2}{E}$ has a well-defined value, and not necessarily that $\ob$ is observed to move between locations $(t,x)$ and $(t+dt,x+vdt)$ in $\fa$. Moreover, the Lorentz transformation \eqref{eq:Lorentz} does not relate coordinates in $\fa$ with their counterpart in $\fb$, rather it relates coordinate displacements $(dt,dx)$ in $\fa$ with corresponding coordinate displacements $(d\t,d\x)$ in $\fb$, once the frames have a well-defined relative velocity.

Importantly, we realise that in the current frame-work the inability to assign the observer $\ob$ simultaneous position and momentum eigenvalues $x\in\RN$ and $p\in\RN$ with reference to $\fa$, is not an inherent feature of this observer. Rather it is an intrinsic feature of the map from the rest-frame of $\oa$ to the rest-frame of $\ob$, and so is more appropriately described as a feature of maps between space-time coordinate systems.

\subsection{Coordinate maps as spinor fields}
While the inverse function theorem as given by equation \eqref{eq:IFT} is expected to be true, its explicit form follows from the  presupposition that space-time should be considered as a four-dimensional manifold (which of course we still hope to accommodate in general). However, as we have seen thus far, the treatment of the relative motion of $\oa$ and $\ob$ may be considered quite generally in  two-dimensional space-time, with any conclusions being extendible to four-dimensional space-time in a relatively simple manner.

In this regard, we continue to investigate the  relative motion of $\oa$ and $\ob$  via a two component map $(t,x)\mapsto(\tb(t,x),\xb(t,x))$. The coordinate transformation from $\fa$ to $\fb$ and its inverse must satisfy the inverse function theorem:
\begin{equation}\label{eq:ift}
\M{p_{0} &  p_{1}  \\ p_{1} & p_{0}  }   \M{p_{0} &  -p_{1}  \\ -p_{1} & p_{0}  }   = \M{m^2c^2 & 0 \\0 & m^2c^2  },
\end{equation}
where $(p_0,p_{1},0,0) = \left(\frac{E}{c},p,0,0\right)$ is the four-momentum of $\ob$ with respect to $\fa$, with $p_{0}=mc\tb_{t}=mc\xb_{x}$ and $p_{1}=-mc^2\tb_{x}=-m\xb_{t}$. Note that equation \eqref{eq:ift} may be reformulated as
\begin{equation}\label{eq:dcom}
 \left[p_{0}\ii_{2}-p_{1}\ps{1}\right]\left[p_{0}\ii_{2}+p_{1}\ps{1}\right]=m^2c^2\ii_{2},
\end{equation}
where $\ii_{2}$ is the $2\times2$-identity matrix and $\ps{1}=\M{0 & 1 \\ 1 & 0}$ is a Pauli spin matrix. It is clear from equation \eqref{eq:QM} that equation \eqref{eq:dcom} may be written in terms of $\psi(t,x)$ according to
\begin{equation}\label{eq:2-comp-wave}
 \left[\p{0}\ii_{2}+\p{1}\ps{1}\right]\left[\p{0}\ii_{2}-\p{1}\ps{1}\right]\psi=-\frac{m^2c^2}{\hbar^{2}}\psi,
\end{equation}
where the right-hand side should be understood to contain a factor of $\ii_{2}$. As such the inverse function theorem may be re-formulated according to
\begin{equation}
 \p{0}^{2}\psi = \left[\left(\p{1}\ps{1}\right)^2-\frac{m^2c^2}{\hbar^2}\right]\psi = \left[\p{1}^{2}-\frac{m^2c^2}{\hbar^2}\right]\psi,
\end{equation}
since $\ps{1}^2=\ii_{2}$. Thus it appears one may identify the differential operator $\p{0}$ with a matrix-differential operator according to
\begin{equation}\label{eq:Hamilton2}
 \p{0}\psi = \left[\alpha\p{1}+i\beta\frac{mc}{\hbar}\right]\psi,
\end{equation}
where $\alpha$ and $\beta$ are $2\times2$-matrices such that
\begin{equation}
 \alpha^2=\beta^2=\ii_{2}\qquad \alpha\beta+\beta\alpha=\mathbb{0}_{2},
\end{equation}
where $\mathbb{0}_{2}$ is the $2\times2$-null matrix.

It is clear that a possible scheme to describe relative motion in four-dimensional space-time is to postulate an extension of \eqref{eq:Hamilton2} according to
\begin{equation}\label{eq:Hamilton4}
 \p{0}\psi = \left[\alpha^{k}\p{k}+i\beta\frac{mc}{\hbar}\right]\psi,\quad k\in\{1,2,3\},
\end{equation}
where the Einstein summation convention is applied to repeated upper and lower indices and
\begin{equation}
 (\alpha^k)^2=\beta^2=\ii\qquad \alpha^{k}\beta+\beta\alpha^{k}=\alpha^{k}\alpha^{l}+\alpha^{l}\alpha^{k}=\mathbb{0}.
\end{equation}
Of course, such a procedure was first developed in Dirac's seminal work on the relativistic theory of the electron (cf. \cite{Dir1928, Dir1981}).
It is common practice to write equation \eqref{eq:Hamilton4} in a more covariant manner according to
\begin{equation}\label{eq:Dirac}
 \left(i\gamma^{\mu}\p{\mu}+\frac{mc}{\hbar}\right)\psi=0,\quad  \gamma^{0}=\beta, \quad \gamma^{k} = - \beta\alpha^{k}.
\end{equation}
The mass-energy constraint now requires four matrices  $\{\gamma^{\mu}\}_{\mu=0}^{4}$  satisfying the anti-commutator relations
\begin{equation}\label{eq:clifford}
 \gamma^{\mu}\gamma^{\nu}+\gamma^{\nu}\gamma^{\mu} = 2\eta^{\mu\nu}\ii,\quad \mu,\nu\in\{0,1,2,3\},
\end{equation}
where $\eta^{\mu\nu}$ are components of the metric tensor $\eta=\diag(1,-1,-1,-1)$.
 It is well established (see \cite{Sch1961} p.71 for instance) that the Clifford algebra generated by \eqref{eq:clifford} has exactly sixteen linearly independent product combinations of the four $\gamma$-matrices, in which case every representation of \eqref{eq:clifford} is realised by a quadruple of $n\times n$ matrices with $n\geq4$.

\section{Relative acceleration}\label{ssec:relativeacceleration}
Differentiating the nonlinear constraint on $\tb$ from equation \eqref{eq:J=1} with respect to $t$ and with respect to $x$ yields
\begin{equation}
 \tb_{t}\tb_{tt}-c^2\tb_{x}\tb_{xt} = 0 \qquad \tb_{t}\tb_{tx}-c^2\tb_{x}\tb_{xx} = 0.
\end{equation}
Multiplying the first of these by $\tb_{t}$ and the second by $c^2\tb_{x}$ and adding the resulting equations, it follows that
\begin{equation}\label{eq:conditiont}
 \left(\tb_{t}^2-c^2\tb_{x}^2\right)\tb_{tt} = 0,
\end{equation}
having also used \eqref{eq:wt}. In a similar manner one may also deduce the condition
\[\left(\xb_{x}^2-\frac{1}{c^2}\xb_{t}^2\right)\xb_{xx} = 0.\]
The condition \eqref{eq:conditiont} now means one of the following conditions must be true:
\begin{equation}\label{eq:mapconditions}
 \tb^{2}_{t}-c^2\tb_{x}^{2} = \frac{1}{mc^2}\left(E^2-p^2c^2\right)=0 \qquad \tb_{tt} = \tb_{xx} = 0,
\end{equation}
in which case the condition \eqref{eq:J=1} fails, or the coordinate map $\ab$ must have components such that $\tb_{t}$ and $\tb_{x}$ are constant, which is only the case when $\oa$ and $\ob$ are in a state of uniform relative motion.

It follows that more general states of relative motion  between the observers which preserve \eqref{eq:enrmom} and the $C^{2}$-regularity of the coordinate map $\ab$ and its inverse cannot be accommodated by the coordinate maps directly. Instead the relative acceleration of $\oa$ and $\ob$ must be accommodated by an external potential $(\phi(t,x),A(t,x))\in C^{2}(\fa)\times C^{2}(\fa)$. It is such potentials which then  account for the variation of the energy-momentum of the observers as they move through the rest-frame of their counterpart.

\subsection{Acceleration via potential fields}\label{sec:potentials}
The observer $\ob$ moving with reference to $\fa$ with unique energy-momentum $(E,p)$, is described by
\begin{equation}
 \tb(t,x) = -i\frac{\hbar}{mc^2}\ln\psi(t,x),
\end{equation}
where $\psi(t,x)$ is a solution of equation \eqref{eq:KG}.
To introduce a relative acceleration between $\oa$ and $\ob$ we introduce a local phase shift in the  coordinate map, of the form $\tb\rightarrow \tb+\frac{1}{mc^2}\Lambda$ where $\Lambda(t,x)\in C^{2}(\fa)$. The \emph{new} coordinate map $\ab$ now has time-component
\begin{equation}\label{eq:phase}
\tb(t, x)+ \frac{1}{mc^2}\Lambda(t,x)=-i\frac{\hbar}{mc^2}\ln\psi(t,x) .
\end{equation}
This corresponds to a local phase-rotation of the function $\psi(t,x)=e^{i(\omega t-kx)}$ according to
\begin{equation}
\psi(t,x)\to e^{i\left(\omega t-kx+\Lambda(t,x)\right)}=e^{\frac{i}{\hbar}\Lambda(t,x)}\psi(t,x).
\end{equation}
The condition $mc^2\tb_{t}=E$ and $-mc^2\tb_{x}=p$ now becomes
\begin{equation}
 \mathcal{E}(t,x)=-i\hbar\frac{\psi_{t}}{\psi} + \Lambda_{t}\quad  \mathcal{P}(t,x)=i\hbar\frac{\psi_{x}}{\psi} - \Lambda_{x},
\end{equation}
which would indicate the space-time-dependence of the energy-momentum of $\ob$ is directly incorporated within the coordinate transformation, which contradicts the conclusions deduced  from equations \eqref{eq:mapconditions}.

Instead, one  introduces a potential field  $(\phi(t,x),A(t,x))\in C^{2}(\fa)\times C^{2}(\fa)$ such that
\begin{equation}\label{eq:potential_eval}
\begin{cases}
  \mathcal{E}\psi=\left(-i\hbar\p{t}+q\phi(t,x)\right)\psi=\left(mc^2\tb_{t}+q\phi(t,x)\right)\psi\\
   \mathcal{P}\psi=\left(i\hbar\p{x}+q A(t,x)\right)\psi=\left(-mc^2\tb_{x}+q A(t,x)\right)\psi,
\end{cases}
\end{equation}
thereby ensuring the $(t,x)$ dependance of $(\mathcal{E},\mathcal{P})$ via $(\phi,A)$.
The constant parameter $q$ represents the coupling strength between the observer $\ob$ and the potential $(\phi,A)$. As such, if $q=0$, the observer $\ob$ does not interact with the potential and so is in a state of uniform relative motion with reference to $\fa$.

If the observer $\ob$ does interact with the potential $(\phi,A)$, then it is still the case that $\ab:\fa\to\fb$ is a map between \emph{inertial reference frames}, even though the observers are accelerating with respect to each other. The principle of general relativity is crucial to this observation since it is this principle which ensures each observer is at rest in its own right thus occupying an inertial reference frame (see \cite{Mol1972} for in-depth discussion of locally inertial coordinate frames).  Since $\tb(t,x)$ is also required to satisfy the wave-equation \eqref{eq:wt}, and the energy-momentum $(\mathcal{E},\mathcal{P})$ is conserved in the sense of equation \eqref{eq:conserve} (cf. \cite{LL2013}) it must be the case the potential  satisfies
\begin{equation}\label{eq:LorentzGauge}
\frac{1}{c^2}\phi_{t} + A_{x} = 0,
\end{equation}
the Lorentz gauge condition.

The constraint $\mathcal{E}^{2}-\mathcal{P}^{2}c^2=m^2c^4$ is imposed since it ensures the rest-energy of $\ob$, namely $mc^2$, is independent of its sate of motion relative to $\oa$, or equivalently with reference to $\fa$. Combined with the eigenvalue problem \eqref{eq:potential_eval}, this generalised energy-momentum constraint now yields the wave-equation
\begin{equation}
 \frac{1}{c^2}\left(-i\hbar\p{t}+q\phi\right)^{2}\psi - \left(i\hbar\p{x}+qA\right)^2\psi - m^2c^2\psi = 0.
\end{equation}
It is clear that this wave equation is naturally extended to relative motion between $\oa$ and $\ob$ in four-dimensional space-time according to
\begin{equation}\label{eq:MinimalCoupling}
 \frac{1}{c^2}\left(-i\hbar\p{t}+q\phi\right)^{2}\psi - \left(i\hbar\nabla+q\mathbf{A}\right)^2\psi - m^2c^2\psi = 0
\end{equation}
with $ \mathbf{A} = (A_{1},A_{2},A_{3})$, which is the Klein-Gordon equation with minimal coupling to a potential field.

\subsection{Relativity of acceleration}
We assign to the observers $\oa$ and $\ob$ the rest-energies and charges $\{Mc^2,Q\}$ and $\{mc^2,q\}$ respectively, which interact via a potential field $(\phi,A)$. Relativity of acceleration is assured if there is a reference frame  $\fa$ wherein $\oa$ has \emph{fixed} energy-momentum $(Mc^2,0)$ and similarly a reference frame $\fb$ where $\ob$ has fixed energy-momentum $(mc^2,0)$.
The conserved energy-momentum of $\ob$ with reference to the frame $\fa$ is given by
 \begin{equation}\label{eq:enr_b}
 (\mathcal{E}(t,x),\mathcal{P}(t,x)) = (E + q\phi(t,x), p+qA(t,x))
 \end{equation}
The energy-momentum of $\oa$ with reference to $\fa$ is
\begin{equation}\label{eq:EpO}
 (\mathcal{E}(t,0),\mathcal{P}(t,0)) = (Mc^2 + Q\phi(t,0), 0+QA(t,0)) = (Mc^2,0)  \quad \forall t\in\RN,
\end{equation}
and so it is required that
\begin{equation}\label{eq:EpO'}
 \phi(t,0)  =  A(t,0) = 0\quad \forall t \in \RN.
\end{equation}
One similarly requires $(\phi'(\t,0),A'(\t,0))=(0,0)$ for all $\t\in\mathds{R}$, where $(\phi',A')$ is the potential field as observed in the reference frame $\fb$.

Given a local phase-shift $\tb(t,x)\to\tb(t,x)+\frac{1}{mc^2}\Lambda(t,x)$, the coordinate maps $\tb(t,x)$ and $\tb(t,x)+\frac{1}{mc^2}\Lambda(t,x)$ satisfy \eqref{eq:MinimalCoupling}, the governing equations describing the motion of $\ob$ with reference to $\fa$,  if the potential field $(\phi,A)$ transforms according to
\begin{equation}\label{eq:gauge}
q\phi \to q\phi -\Lambda_{t} \qquad qA\to qA+\Lambda_{x},
\end{equation}
in which case the eigenvalue problem \eqref{eq:potential_eval} is invariant under local phase shifts of the coordinate map as given by \eqref{eq:phase}. That is to say, if $\psi=e^{\frac{i}{\hbar}mc^2\tb}$, then
\begin{equation}
\begin{aligned}
&\left[\frac{1}{c^2}\left(-i\hbar\p{t}+q\phi-\Lambda_{t}\right)^2 - \left(i\hbar\p{x}+q\phi+\Lambda_{x}\right)^2 - m^2c^2\right] e^{ \frac{i}{\hbar} \Lambda}\psi\\
&\quad =e^{\frac{i}{\hbar}\Lambda}\left[\left(-i\hbar\p{t}+q\phi\right)^2-\left(i\hbar\p{x}+A\phi\right)^2-m^2c^2\right]\psi\\
&\qquad = 0,
\end{aligned}
\end{equation}
where the last equality follows from the fact $\psi$ is a solution of \eqref{eq:MinimalCoupling}. Thus, one may always introduce a local phase shift $\tb\to\tb+\frac{1}{mc^2}\Lambda$, such that
\begin{equation}
(q\phi(t,x),qA(t,x)) \to (q\phi(t,x)-\Lambda_{t}(t,x) , qA(t,x)+\Lambda_{x}(t,x)) = (0,0),
\end{equation}
at a specific $(t,x)$.
The  potential field as observed in the reference frame $\fb$ is now given by
\begin{equation}
\M{\phi'(\tau,\xi)\\ A'(\tau,\xi)} = \frac{1}{mc^2}\M{E & -pc^2 \\-p & E}\M{q\phi(t,x)-\Lambda_{t}(t,x) \\ qA(t,x)+\Lambda_{x}(t,x)} = \M{0\\0}.
\end{equation}
where $(\t,\x)\sim(t,x)$. Of course, the issue remains that one cannot appropriately choose $(t,x)\in\fa$ such that $\ab(t,x)=(\t,0)\in\fb$. Nevertheless, it is clear in principle, there is always a local phase shift $\Lambda(t,x)$ which will ensure \eqref{eq:EpO'} for $\ob$ in $\fb$ also.

Recall that the conservation equation \eqref{eq:conserve} applied to $(\mathcal{E}, \mathcal{P})$ ensured the Lorentz gauge condition \eqref{eq:LorentzGauge}. It is noted that the conservation law \eqref{eq:force} is also expected to be satisfied by $(\mathcal{E}, \mathcal{P})$ and so one also deduces
\begin{equation}\label{eq:phitAx}
(mc^2\tb_{t}+q\phi)_{x} + (-mc^2\tb_{x}+qA)_{t} = 0 \implies \phi_{x}+A_{t} = 0,
\end{equation}
which is clearly gauge invariant under \eqref{eq:gauge} once $\Lambda\in C^{2}(\fa)$. Combining equations \eqref{eq:gauge} and \eqref{eq:LorentzGauge}, one immediately deduces
\begin{equation}
\frac{1}{c^2}\phi_{tt}-\phi_{xx} = 0 \qquad \frac{1}{c^2}A_{tt}-A_{xx} = 0,
\end{equation}
in which case the potential $(\phi,A)$ is a dynamical field whose excitations propagate with velocity $c$ relative to all reference frames.

\section{Conclusion}
The coordinate maps between rest frames of systems of point like observers are an effective method to analyse the relative motion and interactions of these observers. It is found that the structure of the wave-equations and nonlinear constraints applied to these coordinate transformations means the maps may be identified with scalar and spinor wave-functions familiar from relativistic quantum mechanics. The governing equations for these wave-functions may in turn be reduced to the familiar Schr\"{o}dinger equation of non-relativistic quantum mechanics by a number of well-established procedures (see \cite{Gre2000} for instance).

 Moreover, to ensure consistency, it is found that the relative acceleration of the observers cannot be accommodated directly by the coordinate maps, instead it is necessary to introduce potential fields which are found to be familiar massless gauge fields \emph{by requirement}. In the current framework, this gauge redundancy is found to have an essential physical role to play as it ensures each observer may be associated with an inertial coordinate frame throughout the course of the relative acceleration.

\section*{Declaration}
\noindent The author declares there are no conflicts of interest to disclose.

\section*{Funding}
\noindent No funding was received during the preparation of this manuscript.

\section*{Data Availability Statement}
\noindent Data sharing is not applicable to this article as no new data were created or analyzed in this study.


\begin{thebibliography}{10}
\bibitem{Arnetal1999}
M~Arndt, O~Nairz, J~Vos-Andreae, C~Keller, G~Van~der Zouw, and A~Zeilinger.
\newblock Wave--particle duality of $\mathrm{{C}_{60}}$ molecules.
\newblock {\em Nature}, 401:680--682, 1999.

\bibitem{Bre2011}
H~Brezis.
\newblock {\em {Functional Analysis, Sobolev Spaces and Partial Differential
  Equations}}.
\newblock Springer, 2011.

\bibitem{DG1927}
C~Davisson and L~H Germer.
\newblock The scattering of electrons by a single crystal of nickel.
\newblock {\em Nature}, 119:558--560, 1927.

\bibitem{deB1923}
L~de~Broglie.
\newblock Waves and quanta.
\newblock {\em Nature}, 112:540--540, 1923.

\bibitem{deB1925}
L~de~Broglie.
\newblock Recherches sur la theorie des quanta.
\newblock {\em Ann. Phys.}, 10:22--128, 1925.

\bibitem{Die1960}
J~Dieudonne.
\newblock {\em {Foundations of Modern Analysis}}.
\newblock Number~10 in Pure and Applied Mathematics: A Series of Monographs and
  Textbooks. Academic Press, New York, 1960.

\bibitem{Dir1928}
P~A~M Dirac.
\newblock The quantum theory of the electron.
\newblock {\em Proc. Royal Soc. Lond. Ser. A}, 117:610--624, 1928.

\bibitem{Dir1981}
P~A~M Dirac.
\newblock {\em {The Principles of Quantum Mechanics}}.
\newblock Number~27 in Inernational Series of Monographs on Physics. Oxford
  University Press, 1981.

\bibitem{Getal2011}
S~Gerlich, S~Eibenberger, M~Tomandl, S~Nimmrichter, K~Hornberger, P~J Fagan,
  J~T{\"u}xen, M~Mayor, and M~Arndt.
\newblock Quantum interference of large organic molecules.
\newblock {\em Nature Communications}, 2:1--5, 2011.

\bibitem{Gre2000}
W~Greiner.
\newblock {\em Relativistic Quantum Mechanics:Wave Equations}.
\newblock Springer, Berlin, 2000.

\bibitem{LL2013}
L~D Landau and E~M Lifshitz.
\newblock {\em {The Classical Theory of Fields}}, volume~2 of {\em Course of
  Theoretical Physics}.
\newblock Butterworth-Heinemann, 2013.

\bibitem{Mer1998}
E~Merzbacher.
\newblock {\em Quantum Mechanics. Third Edition}.
\newblock Wiley, New York, 1998.

\bibitem{Mol1972}
C~M{\o}ller.
\newblock {\em {The Theory of Relativity}}.
\newblock International Series of Monographs on Physics. Clarendon Press, 1972.

\bibitem{MS1964}
L~Motz and A~Selzer.
\newblock {Quantum mechanics and the relativistic Hamilton-Jacobi equation}.
\newblock {\em Phys. Rev.}, 133:B1622, 1964.

\bibitem{PMWB2019}
J~Pursehouse, A~J Murray, J~W\"atzel, and J~Berakdar.
\newblock Dynamic double-slit experiment in a single atom.
\newblock {\em Phys. Rev. Lett.}, 122:053204, 2019.

\bibitem{Rud1976}
W~Rudin.
\newblock {\em {Principles of Mathematical Analysis}}.
\newblock International Series in Pure and Applied Mathematics. McGraw-Hill New
  York, 1976.

\bibitem{Setal2008}
H~T Schmidt, D~Fischer, Z~Berenyi, C~L Cocke, M~Gudmundsson, N~Haag, H~A~B
  Johansson, A~K{\"a}llberg, S~B Levin, P~Reinhed, et~al.
\newblock Evidence of wave-particle duality for single fast hydrogen atoms.
\newblock {\em Phys. Rev. Lett.}, 101:083201, 2008.

\bibitem{Sch1926}
E~Schr\"{o}dinger.
\newblock Quantisierung als eigenwertproblem.
\newblock {\em Ann. Phys.}, 384:361--376, 1926.

\bibitem{Sch2003c}
E~Schr{\"o}dinger.
\newblock {\em Collected Papers on Wave Mechanics}, volume 302.
\newblock American Mathematical Society, 2003.

\bibitem{Sch1961}
S~S Schweber.
\newblock {\em {An Introduction to Relativistic Quantum Field Theory}}.
\newblock Row, Peterson and Co, Evanston Ill. and Elmsford, NY, 1961.

\bibitem{Setal2020}
A~Shayeghi, P~Rieser, G~Richter, U~Sezer, J~H Rodewald, P~Geyer, T~J Martinez,
  and M~Arndt.
\newblock Matter-wave interference of a native polypeptide.
\newblock {\em Nature Communications}, 11:1--8, 2020.

\bibitem{Wal1984}
R~M Wald.
\newblock {\em {General Relativity}}.
\newblock University of Chicago Press, 1984.

\bibitem{Wig1939}
E~Wigner.
\newblock On unitary representations of the inhomogeneous {L}orentz group.
\newblock {\em Annals of Mathematics}, pages 149--204, 1939.


\bibitem{TM2021}
Tai~Hyun Yoon and Minhaeng Cho.
\newblock Quantitative complementarity of wave-particle duality.
\newblock {\em Science Advances}, 7:eabi9268, 2021.

\end{thebibliography}
\end{document}